\documentclass[11pt, onecolumn, a4paper]{article}
\usepackage{amsmath}
\usepackage{amssymb}
\usepackage{amsthm}
\usepackage{lipsum}
\usepackage{mathptmx}

\usepackage[rm,up,sc,topmarks,calcwidth,pagestyles]{titlesec}
\titleformat{\section}{\Large\bfseries\rmfamily}{\thesection}{1em}{}
\titleformat{\subsection}{\large\bfseries\rmfamily}{\thesubsection}{1em}{}
\titleformat{\subsubsection}{\large\it\rmfamily}{\thesubsubsection}{1em}{}

\usepackage{cite}
\usepackage{algorithmic}
\usepackage[dvipdfmx]{graphicx}
\usepackage{textcomp}

\usepackage{color}

\title{Prediction of Success or Failure for Final Examination using Nearest Neighbor Method to the Trend of Weekly Online Testing}
\author{Hideo Hirose \\
Hiroshima Institute of Technology, Hiroshima, Japan}
\date{}
\begin{document}
\maketitle
\thispagestyle{empty}

\section*{Abstract}
Using the trends of estimated abilities in terms of item response theory for online testing, 
we can predict the success/failure status for the final examination to each student at early stages in courses.
In prediction, we applied the newly developed nearest neighbor method for determining the similarity of learning skill in the trends of estimated abilities, resulting a better prediction accuracy for success or failure.
This paper shows that the use of the learning analytics incorporating the trends for abilities is effective.
ROC curve and recall precision curve are informative to assist the proposed method.
%
%
\\[2mm]
{\it Keywords: }success/failure prediction, item response theory, nearest neighbor, similarity, online testing, learning analytics.




\section{Introduction}

Since a variety of students are now enrolled in universities, it is crucial to identify students at risk for failing courses and/or dropping out as early as possible in order to educate many students altogether, as pointed out in \cite{Siemens2012, Waddington2016}.
However, the more rich in variety, the more we need methodologies for assisting students because conventional methods may not work with the limited number of staffs and classes. New assisting systems shall be required to solve such a difficulty. 

To overcome the difficulty, we established online testing systems aimed at helping students who desire further learning skills for mathematics education.
In such systems, we included the learning check testing, the LCT, for every class to check if students comprehend the contents of lectures or not.
The system has been successfully operating (see \cite{LTLE2016a}, \cite{LTLE2016b}), and some computational results were reported \cite{LTLE2017}. In addition, other relevant cases were well investigated (see \cite{LTLE2016c}, \cite{BIC2016}, \cite{IEE2018}, \cite{PISM2018}, \cite{IJSCAI2017}, \cite{LTLE2016d}). 

As indicated in \cite{Elouazizi}, \cite{Siemens2015}, and \cite{Siemens2012}, the immediate research attention for learning analytics is crucial to make a sustainable impact on the research and practice of learning and teaching. Using the outputs obtained from the online testing, it is not so difficult to collect a large-scale of learning data.
We may be able to actively tackle the collected data to find the optimal strategies for better learning methods. It is also important to analyze the data theoretically (see \cite{WiseShaffer}). 

This paper is aimed at obtaining effective learning strategies for students at risk for failing courses and/or dropping out, using a large-scale of learning data collected from the online testings.
In this paper, unlike the conventional methods using the correct answer rate (CAR) to identify the ability of a student (e.g., see \cite{LTLE2017}), we use the ability obtained from the item response theory (IRT, e.g., see \cite{Ayala}, \cite{Hambleton91}, \cite{LindenHambleton}), and we show a new method to identify students at risk as early as possible using the IRT results. 


\section{Weekly Online Testing}

Analysis basic (i.e., calculus) and linear algebra are two fundamental subjects that mathematics teachers are involved in the weekly online testings. 
Testing time duration is ten minutes, and $m$ questions using multiple choice are provided to each testing; in 2017 semesters, $m=5$. 
The testings to check the comprehension of each unit are incorporated into regular classes; for example, in the case of analysis basic, {\it differentiation} unit has a set of question items for testing, and calculus online testings consist of 14 different such sets.
Each subject (analysis basic or linear algebra) consists of 16 units including midterm and final examinations; except for two examination classes, students have 14 lectures incorporating the LCT; if we denote $K$ as the number of opportunities that students take the LCT, $K=14$ in 2017 semesters. 
In addition, we define the number of freshman students to be enrolled as $N$;  in 2017 semesters, $N$ is approximately 1,100. 
Thus, we have user-item response matrices sized of $N \times mK$ to each subject at the end of the semesters.

Figure 1 shows a part of such a response matrix; row and column correspond to student id and item (question) id; a red color element indicates that a student solved a problem item successfully, and a green color element means to be a failed response.


\begin{figure}[htbp]
\begin{center}
\includegraphics[height=6cm]{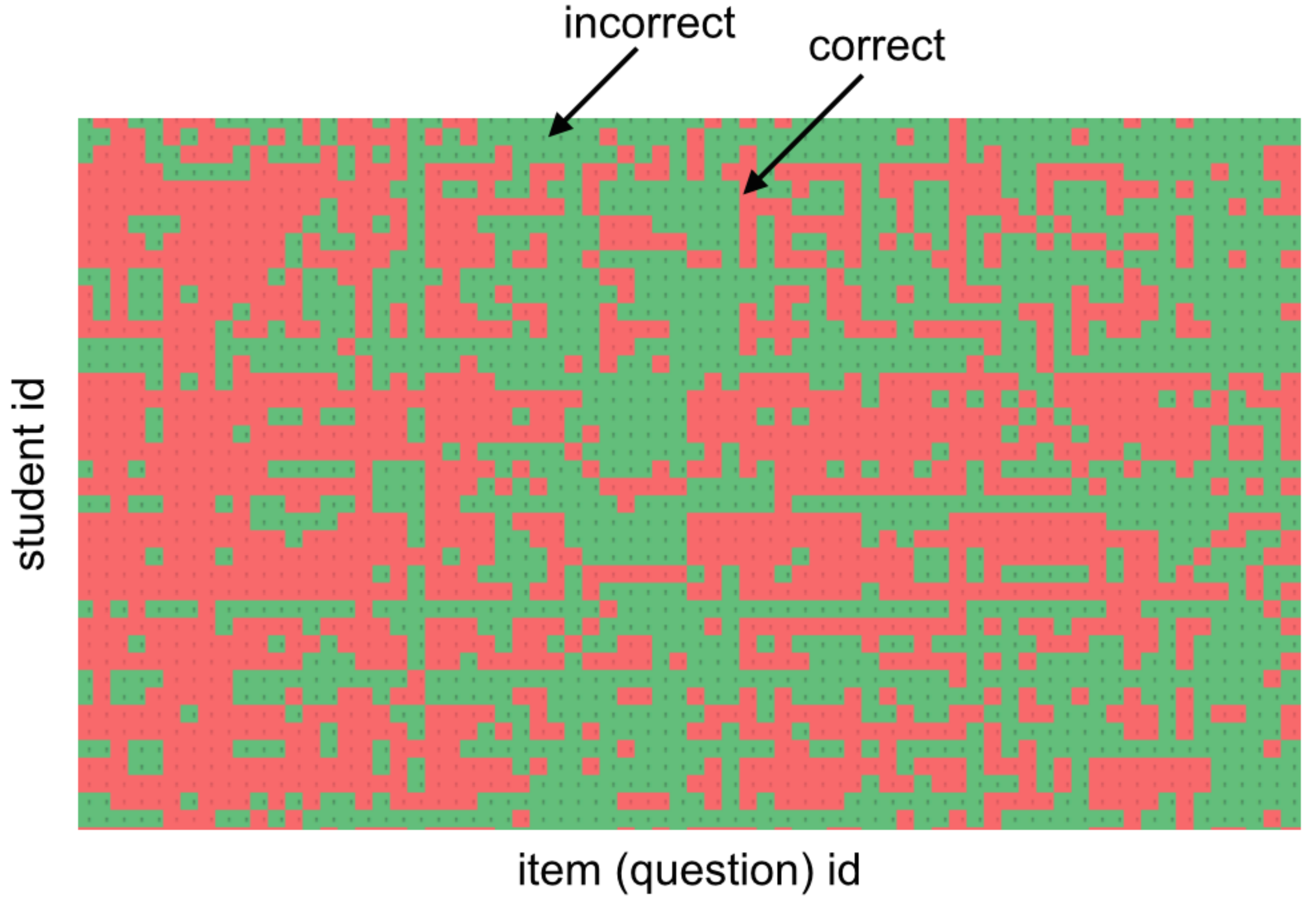}
\end{center}
\caption{A part of a item response matrix (analysis basic in the first semester in 2017).}
\end{figure}

\section{Ability Evaluation Using the IRT}

In many cases, evaluation for learning skill is assessed by using the correct answer rate (CAR) to questions; CAR values are obtained by the ratio of the number of correct answers to the given questions. Although this criterion is easily understood, it does not include the effects from other students' scores.

The item response theory (IRT) provides us the difficulties of the test items (problems) and the examinees' abilities together, 
resulting in evaluating the examinees' abilities accurately and fairly. In addition, adaptive testing using the IRT selects the most appropriate items to examinees automatically, resulting in more accurate ability estimation and more efficient test procedures (see \cite{CATE}, \cite{TALE2012}, \cite{TALE2014a}, \cite{TALE2014b}, \cite{TALE2014c}, \cite{IPSJ2014}). 
Thus, we incorporated the IRT evaluation method into the online testing systems. In this paper, we deal with the cases of the standard IRT evaluation using 
the two-parameter logistic function $P(\theta_i;a_j,b_j)$ shown below. 
\begin{eqnarray} \nonumber
P(\theta_i;a_j,b_j)&=&{1 \over 1+\exp\{-1.7a_j(\theta_i-b_j)\} },\\
&=&1-Q(\theta_i;a_j,b_j),
\end{eqnarray}
where $\theta_i$ expresses the ability for student $i$, and $a_j, b_j$ are constants in the logistic function for item $j$ called the discrimination parameter and the difficulty parameter, respectively.
The corresponding likelihood function for all the examinees, $i=1,2,\dots,N$, and all the items, $j=1,2,\dots,n$, will become
\begin{eqnarray} 
L=\prod_{i=1}^N \prod_{j=1}^n \left(P(\theta_i;a_j,b_j)^{\delta_{i,j}} 
\times Q(\theta_i;a_j,b_j)^{1-\delta_{i,j}} \right),
\end{eqnarray}
where $\delta_{i,j}$ denotes the indicator function such that $\delta=1$ for success and $\delta=0$ for failure in answering a question. 
We adopt the IRT evaluation for students' abilities unlike the case in \cite{LTLE2017}.



\section{Trend of Estimated Students' Abilities Using Each Unit Response Matrix in the IRT}

First, we show some trends for the estimated abilities to each unit. This means that we use the response matrices $M_k(N,m), \ k=1,\dots, K$. 
Then, we define $\theta_{0}(i,k)$ as student $i$'s ability using the $k$th LCT response results, where each response matrix is a $N \times m$ size matrix. 

Figure 2 shows a part of such a case for analysis basic; in this demonstration, $N$ is about 100, and this corresponds to students for some department.
The figure indicates that it seems difficult to discriminate students into certain categories. We see many up-and-down ability estimates in the ability trends from the 1st LCT to 14th LCT. The small number of question items may make the variance of the estimates large. That is, the ability estimates using each LCT response matrix are unreliable. 

\begin{figure}[htbp]
\begin{center}
\includegraphics[height=7.5cm]{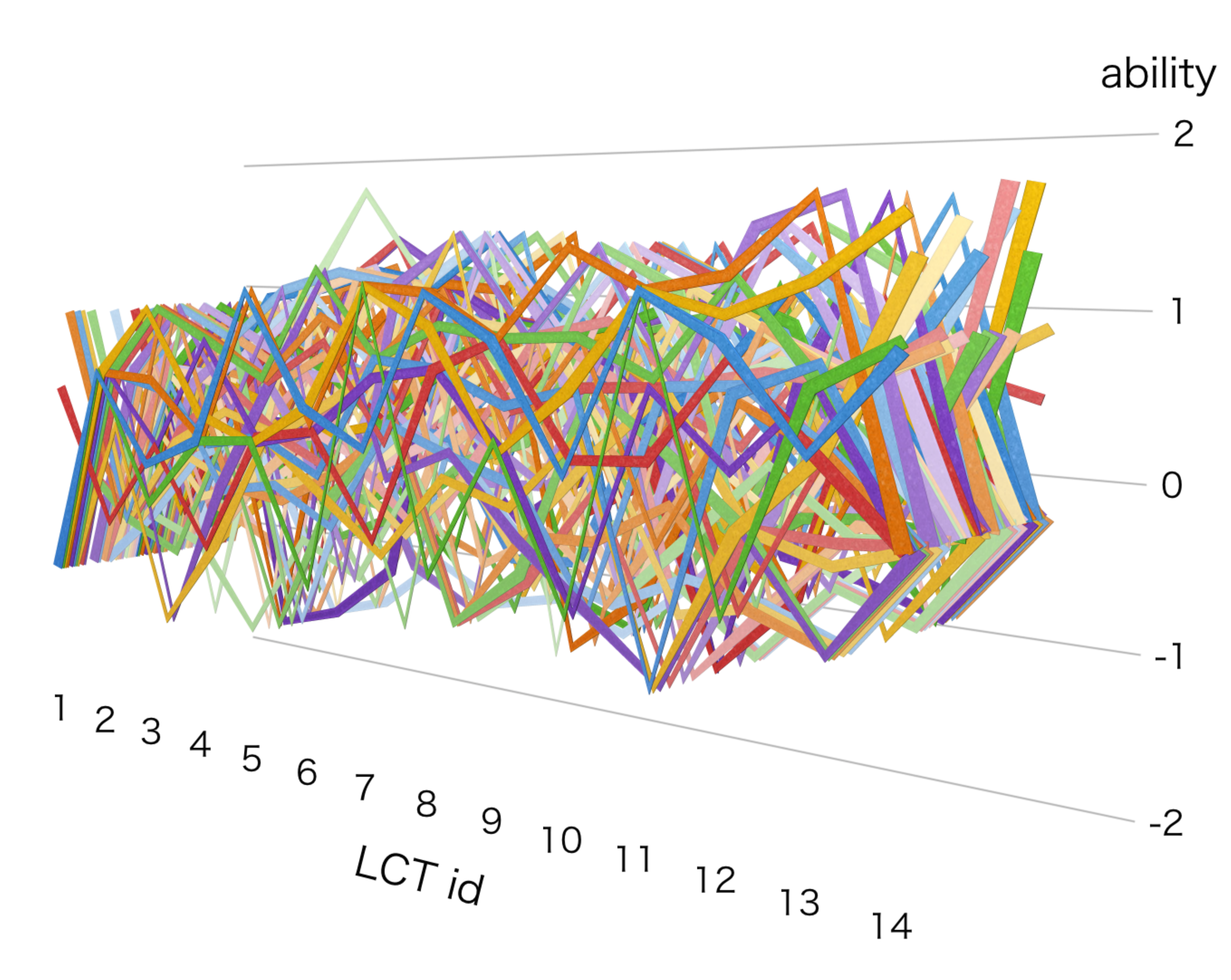}
\end{center}
\caption{Some trends for the estimated abilities $\theta_{0}(i,k)$ to each unit (analysis basic in the first semester in 2017).}
\end{figure}

We can also see that mean trends of abilities to each student show a slight ascending tendency. However, this is actually resulting from ascending difficulties as lectures go forward, i.e., the more lectures students take, the more difficult the lecture level becomes. Thus, this tendency could be ignored.

\section{Identifying Successful/Failed Students Using the Full Response Matrix in the IRT}

To identify students at risk, the use of known two categorized groups could be helpful: one is the successful students for the final examination, and the other is the failed students.
Figure 3 shows the histogram of estimated abilities of LCT to successful students overlaid the histogram of estimated abilities of LCT to failed students  in the case of analysis basic in the first semester in 2017.
The numbers of successful students is 921, and failed students is 206; the ratio of failed students to all the students is $0.18$.
Here, we have used full response matrices $M(N,mK)$ in the estimation to obtain the most reliable estimates for abilities.

Except for very low values of ability estimates, the histograms indicate the normal distribution with different mean values (around $0.22$ for successful students and $-0.57$ for failed students); the lowest estimates around $-3.0$ in both groups were resulting from the absence for testing. However, it seems very difficult to discriminate students into two groups by using certain ability threshold value. When we adopt the decision tree method, the most appropriate ability threshold values becomes to be $-0.047$.

\begin{figure}[htbp]
\begin{center}
\includegraphics[height=6cm]{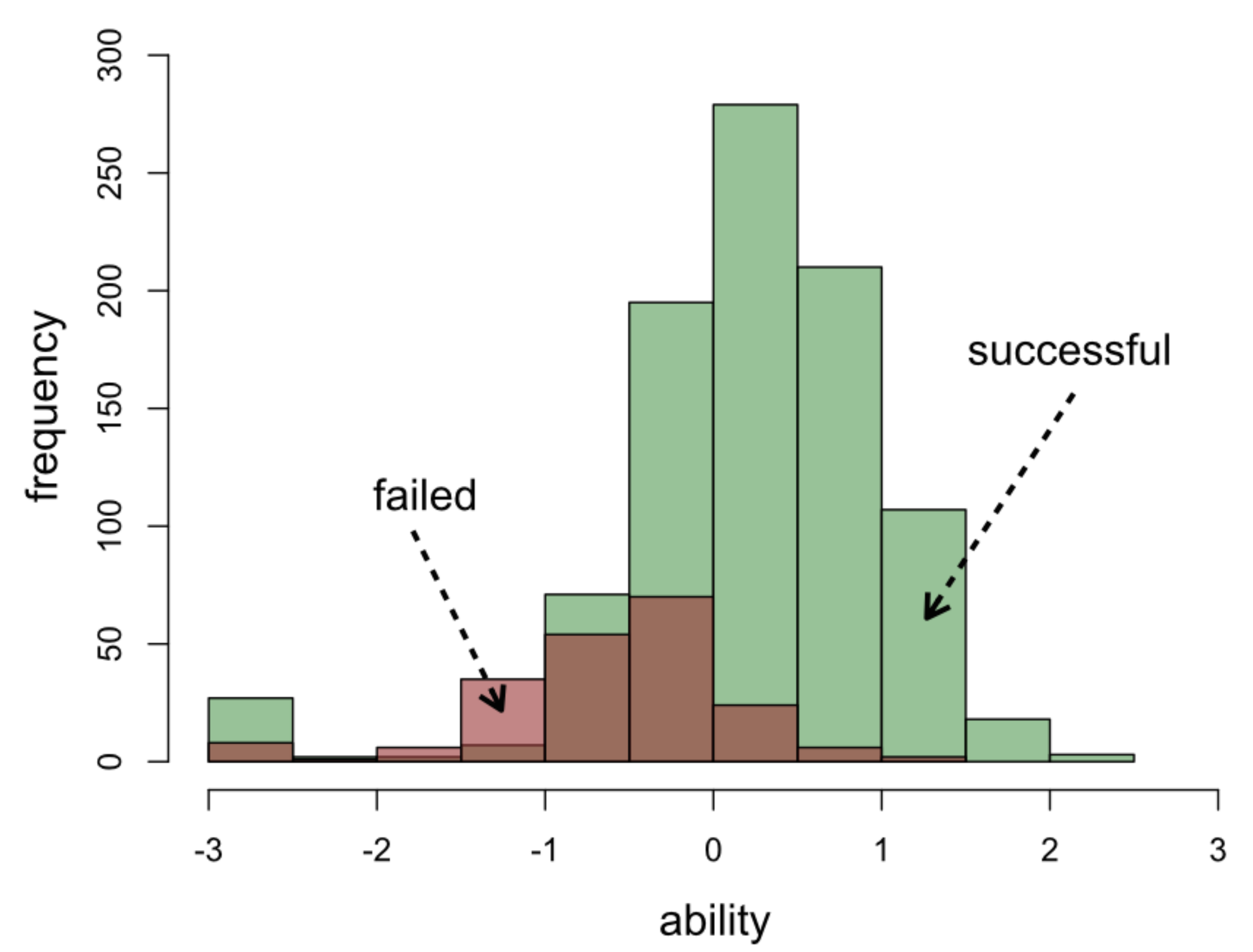}
\end{center}
\caption{Histograms of estimated abilities for successful/failed two groups (analysis basic in the first semester in 2017).}
\end{figure}

The confusion matrix using this threshold is illustrated in table 1. The misclassification rate for this confusion matrix is $0.28$.  Limited to failed students, the decision tree predicted 446 students may fail, and 169 students actually failed; the hitting ratio is $38\%$, and the result seems not to be useful.

\begin{table}[htbp]
\caption{Confusion matrix determined by decision tree using full response matrix.}
\begin{center}
\begin{tabular}{cc|ccc}
\hline
&&&predicted \\
 &  & successful & failed & total \\
\hline
                 & successful & 644 & 277& 921 \\
observed  & failed & 37 & 169 & 206\\
                 & total & 681 & 446 & 1127\\
\hline
\multicolumn{4}{l}{\qquad \qquad \qquad \qquad \quad threshold $= -0.047$}
\end{tabular}
\label{tab1}
\end{center}
\end{table}


In addition to the LCT results, we have incorporated the placement test (PT) results taken at the very beginning of the first semester. We have two kinds of tests: one is rather fundamental test and the other is advanced test in high school level. Using the fundamental PT and the LCT results, we plotted the correlations for these two tests in three groups in Figure 4 in the case of analysis basic in the first semester in 2017: first group is the successful in the final examination (score range is 60-100 expressed by green dots in the figure), second group is the badly failed group (score range is 0-39 expressed by red dots), and the rest is the group (score range is 40-59 expressed by yellow dots). The horizontal axis means the ability values standardized to the standard normal distribution, and the vertical axis means the fundamental PT score. 
Although the information using the computational results via the IRT is added, it is still hard to find the boundaries to classify students into three groups or two successful/failed groups.
In order to discriminate successful students from failed students much more clearly, it would be recommended to include other kind of information.

\begin{figure}[htbp]
\begin{center}
\includegraphics[height=5.5cm]{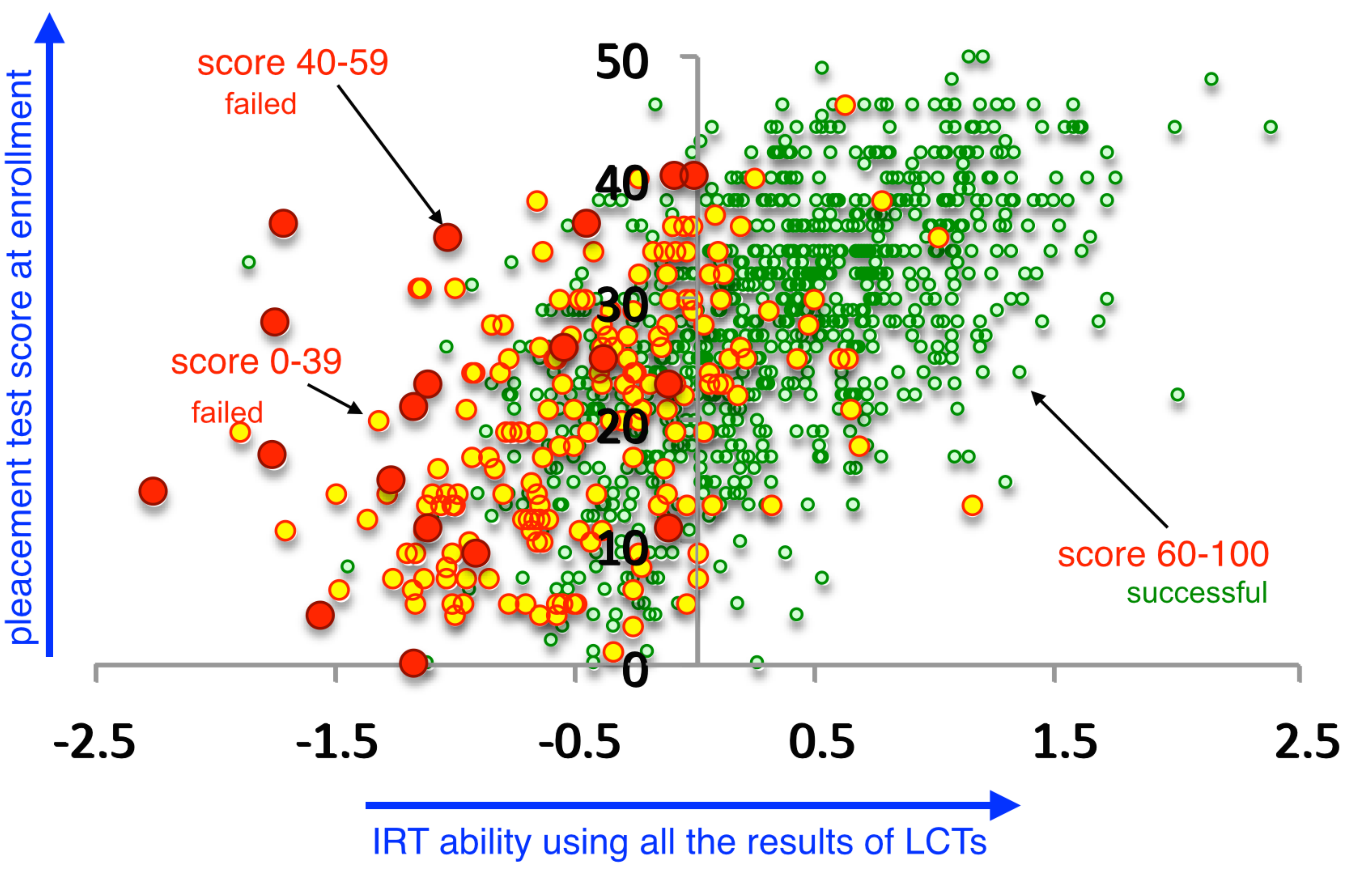}
\end{center}
\caption{Correlations for the LCT results and the placement test results in three successful/failed groups (analysis basic in the first semester in 2017).}
\end{figure}

\section{Trend of Estimated Students' Abilities Using Full Units Response Matrix in the IRT}

We define $\theta_{1}(i,k)$ as student $i$'s ability using the response results from the 1st LCT to $k$th LCT, that is, the response matrix becomes a $N \times km$ size matrix. 
Figures 5 and 6 show the trends of estimated abilities $\theta_{1}(i,k)$ for successful and failed groups. Looking at Figure 5, we can see that the estimated ability to each student seems to converge to a certain value as lectures go forward, and this means that the estimates become accurate. Figure 6 tells us that the estimated abilities show rather small variations around 0 value initially, but later they become lower as lectures go forward. 
Comparing to Figure 2, Figures 5 and 6 seem to characterize the trends of estimated abilities for two groups with higher reliability than Figure 2 seems to. However, how can we use such a vague trend tendency to categorize the student groups into successful/failed students? We have to develop some tools to measure the similarity of trend numerically.


\begin{figure}[htbp]
\begin{center}
\includegraphics[height=7.5cm]{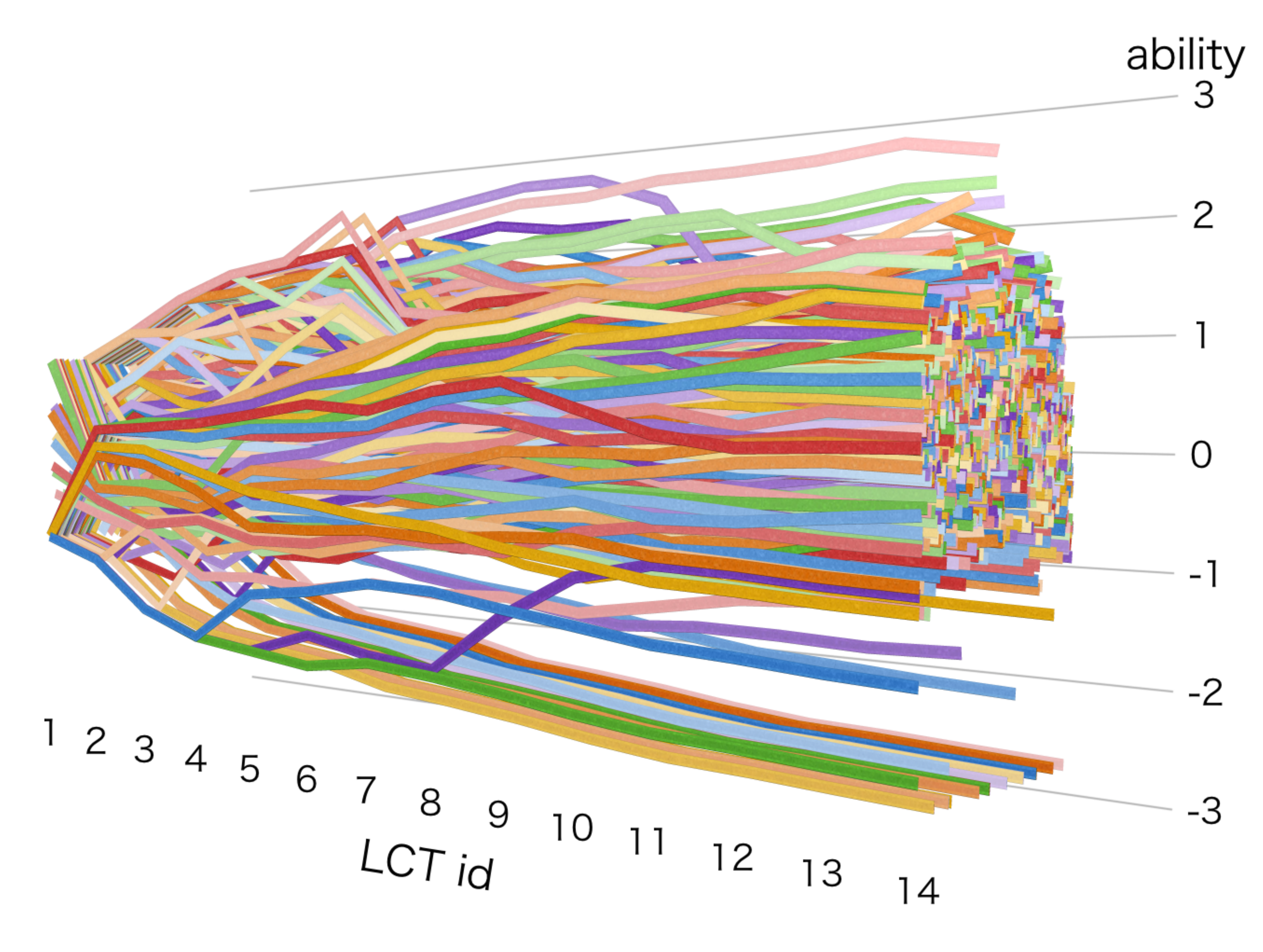}
\end{center}
\caption{Trends of estimated abilities $\theta_{1}(i,k)$ for successful group (analysis basic in the first semester in 2017).}
\end{figure}

\begin{figure}[htbp]
\begin{center}
\includegraphics[height=7.5cm]{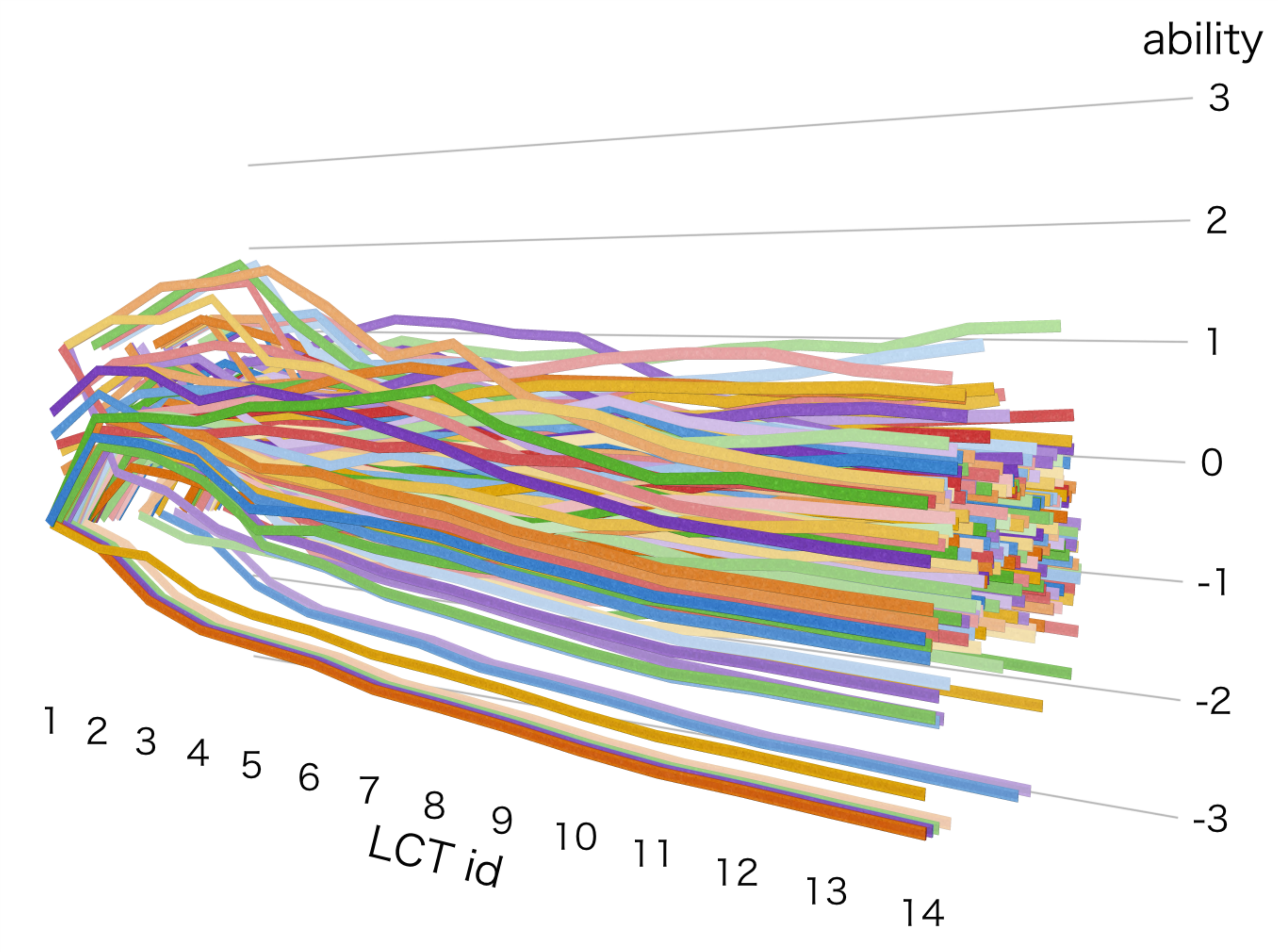}
\end{center}
\caption{Trends of estimated abilities $\theta_{1}(i,k)$ for failed group (analysis basic in the first semester in 2017).}
\end{figure}

\section{Similarity Identification by Nearest Neighbor}

In order to identify successful/failed students with much higher reliability in prediction, we here define the similarity via the nearest neighbor using the estimated ability trends as lectures goes forward.  
To do this, we use $\theta_{1}(i,k)$ defined in the previous section by incorporating the tentative response matrices $M_{m,k}(N,mk), \ k=1,\dots, K$ using LCT no.1 to no.$k$.

As an example to explain the similarity, we have provided Figure 7, where we can see three students' ability trends using estimated abilities from LCT no.1 to no.7. As lectures go forward, the estimated abilities seem to tend to certain values although the values are unreliable at the early stages of the trends.
We may assume that two final destination of success/failure may be the same if the estimated trends for abilities are close to each other. Although the use of only the full response matrices $M_{m,K}(N,mK)$ did not bear a reliable results, we may expect that the trends of ability estimates by using response matrices $M_{m,k}(N,mk), \ k=1,\dots, K$ will give us much more information. 

\begin{figure}[htbp]
\begin{center}
\includegraphics[height=5.5cm]{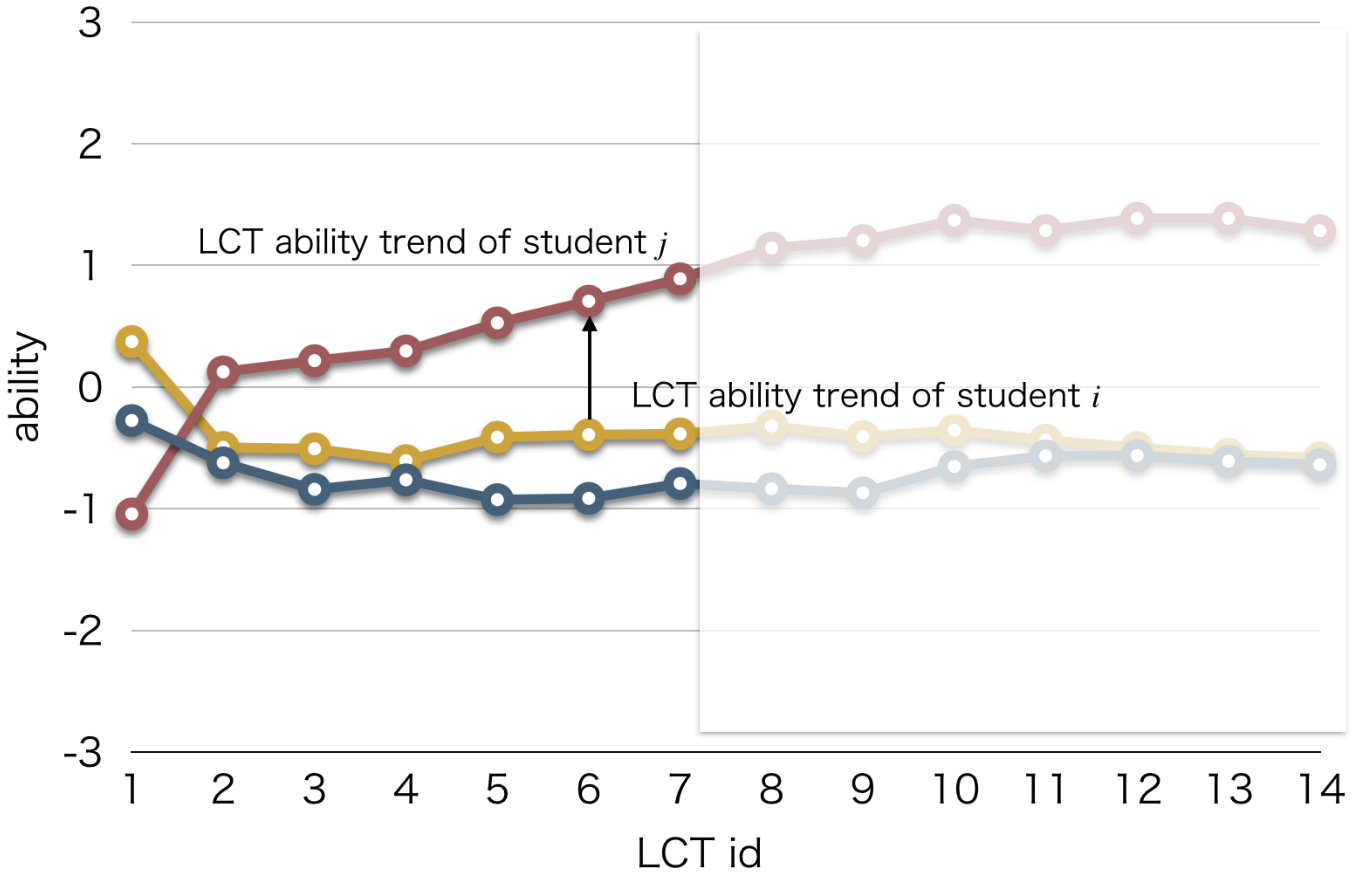}
\end{center}
\caption{An example to explain the similarity via the nearest neighbor using the estimated ability trends using no.1 to no.$7$ LCTs.}
\end{figure}

We define the similarity of the two ability trends ($i$ and $j$) by the following formula
$S_{i,j}^k$ such that
\begin{eqnarray} 
S_{i,j}^k=\sqrt {{1 \over k} \sum_{l=1}^k (\theta_{1}(j,l)-\theta_{1}(i,l))^2 }, \ (i \ne j),
\end{eqnarray}
then, we can consider that $S_{i,j}^k$ expresses the mean distance between the trends of abilities for students $i$ and $j$ from the 1st LCT to $k$th LCT.

Sorting $S_{i,j}^k$ in ascending order in terms of $j$ such as $S_{i,(1)}^k \le \dots, \le S_{i,(N-1)}^k$, 
$S_{i,(j)}^k$ expresses the ordered statistics of $\{ S_{i,j}^k \}$.
We select the 10 least $S_{i,(j)}^k$ (i.e., $S_{i,(1)}^k, \dots, S_{i,(10)}^k$), and obtain the mean value $\mu(i,k)$ of these final examination's success/failure indicator functions $\delta_{i,(j)}^k$, i.e., 1 for success and 0 for failure from $(j)$th final success/failure results. Then, $\mu(i,k)=0, 0.1, \dots, 0.9, 1$, and we can consider that $\mu(i,k)$ expresses the predicted value for success in the final examination. We next show the investigation results on the prediction accuracy using this similarity definition.

\section{Identifying Successful/Failed Students Using the Similarity of the Trends of Estimated Students' Abilities in the IRT}

We consider typical three cases in using the LCT response results: 1) from LCT no.1 to LCT no.4, 2) from LCT no.1 to LCT no.7, 3) from LCT no.1 to LCT no.11. 

Figure 8 shows the bar charts for the predicted number of students to be failed in the final examination in the case of analysis basic in the first semester in 2017.
Upper green parts express the observed successful number of students; lower orange parts express the observed failed number of students. In the figure, we see a notation of $p \ge 0.3$, e.g., which is the same as $\mu(i,4) \ge 0.3$ when using LCT no.1 to LCT no.4, and other notations are expressed in a similar manner.
For example, in the case of 2) from LCT no.1 to LCT no.7, and $p \ge 0.4$, we predicted that 173 students are to be failed in which 69 students are actually failed and 104 students are actually successful. These numbers are also seen in table 2.

Although the observed failed number of students 206 is larger than the predicted value, the hitting ratio, 0.40, shows larger value to some extent than that shown in section 4 where the size of the response matrix is the maximum. 
Looking at all the bar charts in the figure, it should be noted that all the three cases using LCT no.1 to no.4, no.1 to no.7, and no.1 to no.11 reveal that the hitting ratios are larger than that shown in section 4 as long as $p \ge 0.4$. 

From the confusion matrix in table 2, we can easily obtain the misclassification rates as shown in table 3.
For example, to the cases $p \ge 0.3$, $p \ge 0.4$, and $p \ge 0.5$ using LCT no.1 to no.7 which used almost half of the LCT, the misclassification rates are 0.28, 0.22, 0.18. All the misclassification rates in table 3 are smaller than or equal to that computed in section 5 which used all the LCT results in computing the IRT abilities.

\begin{figure}[htbp]
\begin{center}
\includegraphics[height=5cm]{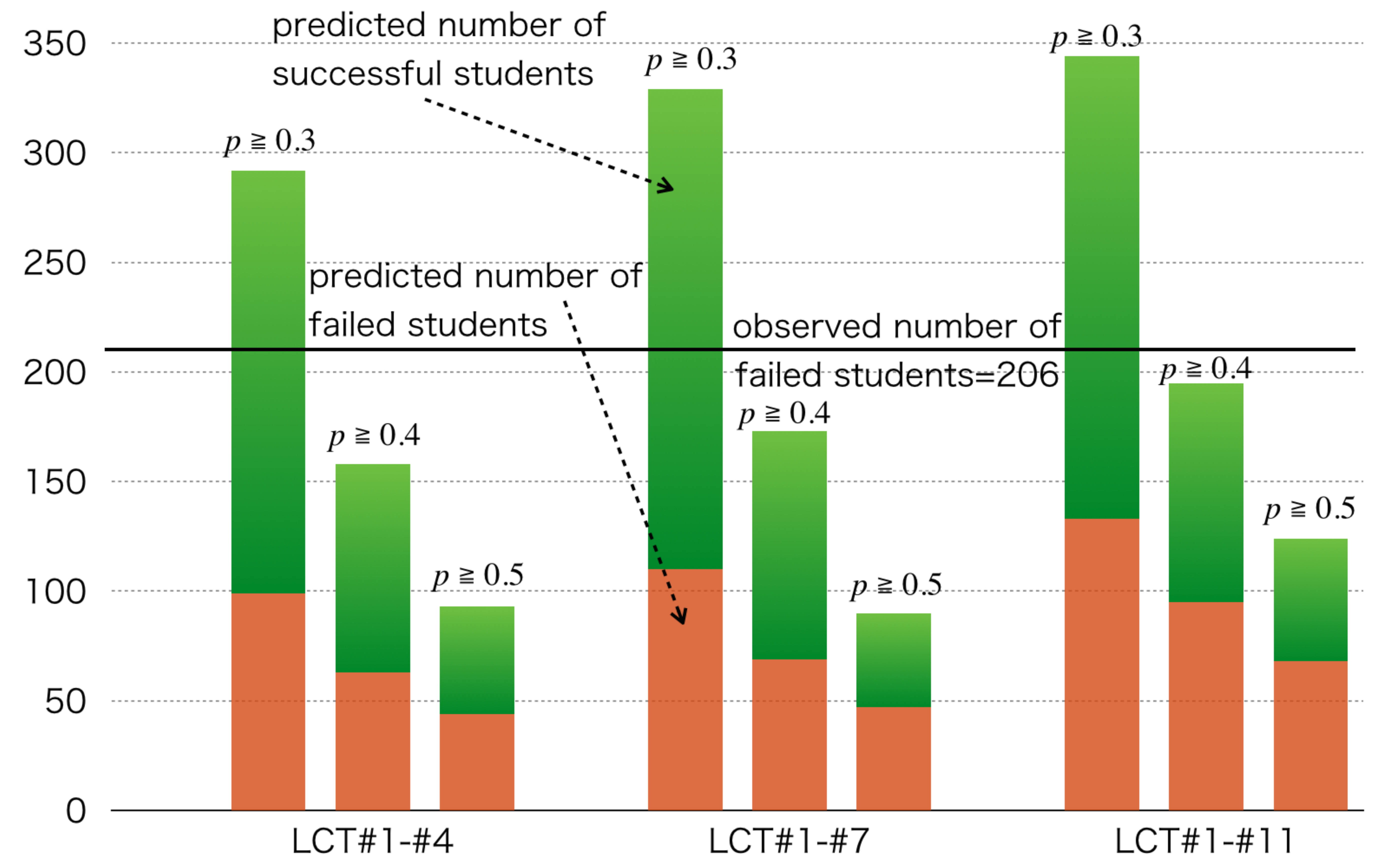}
\end{center}
\caption{Numbers of successful/failed students using the similarity of the trends of estimated students' abilities (analysis basic in the first semester).}
\end{figure}

\begin{table}[htbp]
\caption{Confusion matrix determined by the nearest neighbor (analysis basic)}
\begin{center}
\begin{tabular}{cc|ccc}
\hline
LCT \#1-\#4&$p \ge 0.3$&&predicted \\
 &  & successful & failed & total \\
\hline
                 & successful & 728 & 193& 921 \\
observed  & failed & 107 & 99 & 206\\
                 & total & 835 & 292 & 1127\\
\hline 
\hline
LCT \#1-\#4&$p \ge 0.4$&&predicted \\
 &  & successful & failed & total \\
\hline
                 & successful & 826 & 95& 921 \\
observed  & failed & 143 & 63 & 206\\
                 & total & 969 & 158 & 1127\\
\hline 
\hline
LCT \#1-\#4&$p \ge 0.5$&&predicted \\
 &  & successful & failed & total \\
\hline
                 & successful & 872 & 49& 921 \\
observed  & failed & 162 & 44 & 206\\
                 & total & 1034 & 93 & 1127\\
\hline
\hline
LCT \#1-\#7&$p \ge 0.3$&&predicted \\
 &  & successful & failed & total \\
\hline
                 & successful & 702 & 219& 921 \\
observed  & failed & 96 & 110 & 206\\
                 & total & 798 & 329 & 1127\\
\hline 
\hline
LCT \#1-\#7&$p \ge 0.4$&&predicted \\
 &  & successful & failed & total \\
\hline
                 & successful & 817 & 104& 921 \\
observed  & failed & 137 & 69 & 206\\
                 & total & 954 & 173 & 1127\\
\hline 
\hline
LCT \#1-\#7&$p \ge 0.5$&&predicted \\
 &  & successful & failed & total \\
\hline
                 & successful & 878 & 43& 921 \\
observed  & failed & 159 & 47 & 206\\
                 & total & 1037 & 90 & 1127\\
\hline
\hline
LCT \#1-\#11&$p \ge 0.3$&&predicted \\
 &  & successful & failed & total \\
\hline
                 & successful & 710 & 211& 921 \\
observed  & failed & 73 & 133 & 206\\
                 & total & 783 & 344 & 1127\\
\hline 
\hline
LCT \#1-\#11&$p \ge 0.4$&&predicted \\
 &  & successful & failed & total \\
\hline
                 & successful & 821 & 100& 921 \\
observed  & failed & 111 & 95 & 206\\
                 & total & 932 & 195 & 1127\\
\hline 
\hline
LCT \#1-\#11&$p \ge 0.5$&&predicted \\
 &  & successful & failed & total \\
\hline
                 & successful & 865 & 56& 921 \\
observed  & failed & 138 & 68 & 206\\
                 & total & 1003 & 124 & 1127\\
\hline
\end{tabular}
\label{tab1}
\end{center}
\end{table}

\begin{table}[htbp]
\caption{Misclassification rates by using the decision tree  (analysis basic)}
\begin{center}
\begin{tabular}{c|ccc}
\hline
&LCT \#1-\#4&LCT \#1-\#7&LCT \#1-\#11 \\
\hline
$p \ge 0.3$ & 0.27 & 0.28 & 0.25 \\
$p \ge 0.4$ & 0.21 & 0.22 & 0.19\\
$p \ge 0.5$ & 0.19 & 0.18 & 0.17\\
\hline
\end{tabular}
\label{tab1}
\end{center}
\end{table}


Table 4 shows the hitting ratios of the number of actually failed students to the number of predicted failed students corresponding to table 2. Since the hitting ratio using all the LCT results was 0.38 as mentioned in section 5, the hitting ratios using the nearest neighbor similarity are superior to that using the IRT abilities from all the LCT results.

\begin{table}[htbp]
\caption{Hitting ratios of the number of actually failed students to the number of predicted failed students (analysis basic)}
\begin{center}
\begin{tabular}{c|ccc}
\hline
&LCT \#1-\#4&LCT \#1-\#7&LCT \#1-\#11 \\
\hline
$p \ge 0.3$ & 0.34 & 0.33 & 0.39 \\
$p \ge 0.4$ & 0.40 & 0.40 & 0.49\\
$p \ge 0.5$ & 0.47 & 0.52 & 0.55\\
\hline
\end{tabular}
\label{tab1}
\end{center}
\end{table}

\section{Discussions}

Comparing to the misclassification rate in the condition that only the numbers of success/failures are known, the predicted misclassification rates seem not to be informative so much. That is, in the analysis basic case the success rate is 0.82 and the failure rate is 0.18, then misclassification rate will be 0.18 if we assume that all the students are successful in the final examination. The estimated misclassification rates in table 3 are comparative at most or worse than that in the case mentioned above. 
However, it is totally absurd that we admit all the students are successful; we cannot find any students at risk. The hitting ratio is 0.

We actually want to know the students at risk, and the important point is that we can find such students with high probability. From this viewpoint, the high hitting rates are informative to tell such students that you may fail if you insist to continue the same behavior. In the analysis basic case, 18\% students failed, and we could point out about half of such students. 


Using the obtained value of $p$ which means the estimated failure probability using the trend of accumulated IRT results, we will be able to make alert to students for possible failures in the coming final examination. One method is to use the estimated value directly such that ``you will fail in the final examination with probability of $p$". However, it seems that two-value information of failure or success is much clearer to students such that ``you will fail in the final examination as long as you leave your learning style unchanged".

In such a situation, the threshold values for $p$ will be informative when we alert students to the signal for possible failures. ROC (receiver operating characteristic) curve may help to find such a threshold value. 
Figure 9 shows ROC curves when we use from LCT no.1 to LCT no.4, from LCT no.1 to LCT no.7, and from LCT no.1 to LCT no.11, in the case of analysis basic in the first semester. 
When we abbreviate false positive rate and true positive rate to FPR = FP / (FP + TN) and TPR = TP / (TP + FN), respectively, ROC curve represents the relationship between FPR and TPR, where FP, TN, TP, and FN are false positive, true negative, true positive, and false negative, respectively.
In the figure, false positive rate in the abscissa means the ratio of the number of actually successful students in the predicted failed students to the total number of actually successful students, and true positive rate in the ordinate means the ratio of the number of actually failed students in the predicted failed students to the total number of actually failed students.
In Figure 9, we see that $0.2 \le p \le 0.4$ could be used for the threshold value. However, from the viewpoint of the importance of true positive rather than false positive, we recommend to use the case of $p = 0.4$ in this case. We paid attention much to true positive, i.e., students at risk.

To understand the hitting ratio shown in table 4, recall precision curve may be useful.
Figure 10 shows the recall precision curves we use from LCT no.1 to LCT no.4, from LCT no.1 to LCT no.7, and from LCT no.1 to LCT no.11, in the case of analysis basic in the first semester. 
Recall and precision mean TP / (TP + FN) and TP / (TP + FP), respectively, and recall precision curve represents these two relationship.
In the figure, recall in the abscissa means the ratio of the number of actually failed students in the predicted failed students to 
the total number of actually failed students, and recision in the ordinate means the ratio of the number of actually failed students in the predicted failed students to the total number of predicted failed students.
Precision is equivalent to hitting ratio shown in table 4. 
In table 4, the hitting ratio is $38$, and we see that this value is lower than those of $0.40, 0.40, 0.49$ for $p \ge 0.4$ in LCT no.1 to LCT no.4, LCT no.1 to LCT no.7, and LCT no.1 to LCT no.11 cases.

Figure 11 shows the predicted numbers of successful students and failed students to each $p$ using results from LCT no.1 to LCT no.11 in the case of analysis basic in the first semester.
We can see that the number of predicted successful students are becoming smaller when $p$ is larger than 0.4.

%
%
%

\begin{figure}[htbp]
\begin{center}
\includegraphics[height=7cm]{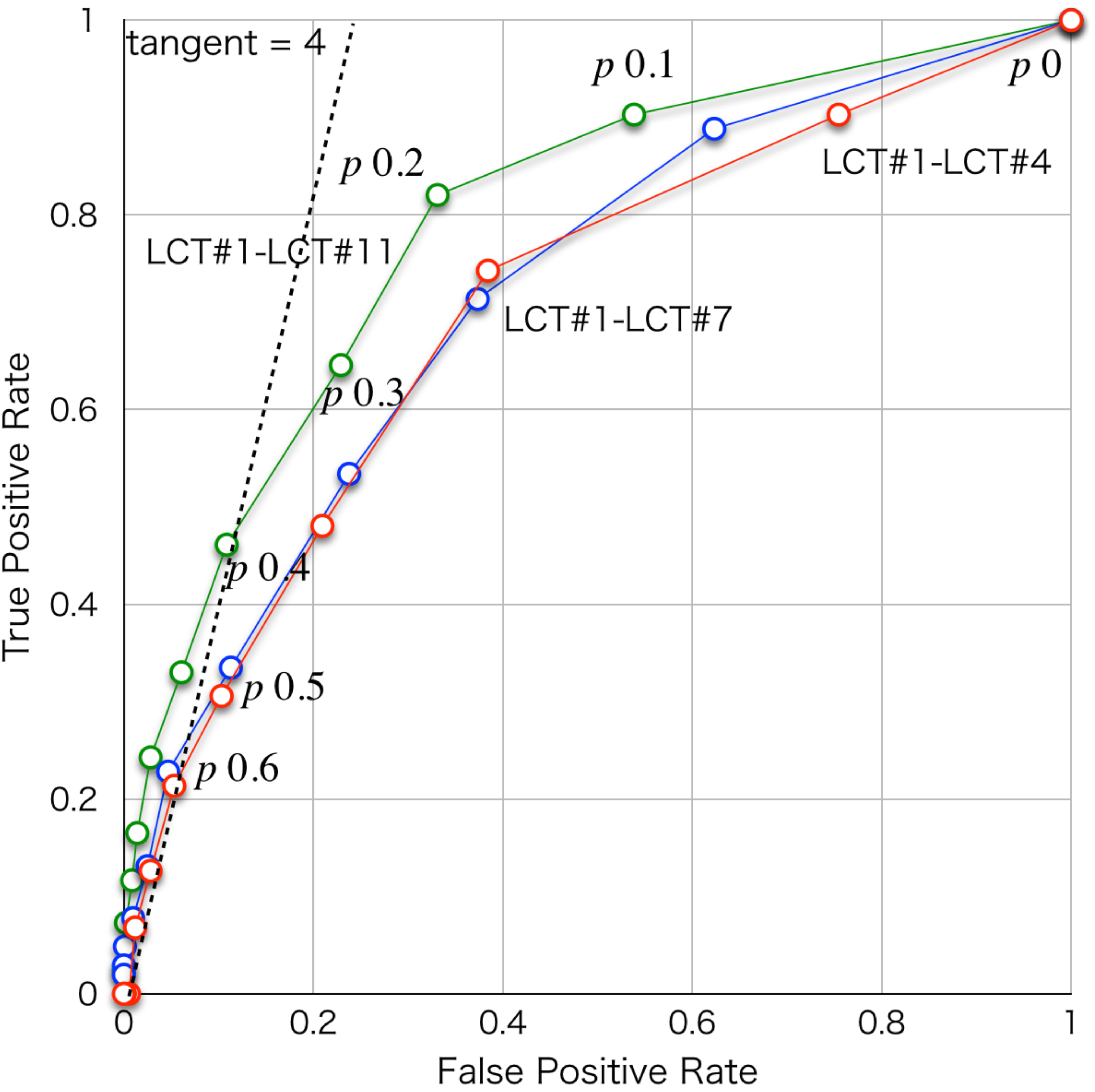}
\end{center}
\caption{ROC curve (analysis basic in the first semester). 
False Positive Rate in the abscissa means 
the ratio of 
the number of actually failed students in the predicted failed students 
to 
the total number of actually successful students. 
True Positive Rate in the ordinate means 
the ratio of 
the number of actually failed students in the predicted failed students 
to 
the total number of actually failed students.}
\end{figure}

\begin{figure}[htbp]
\begin{center}
\includegraphics[height=7cm]{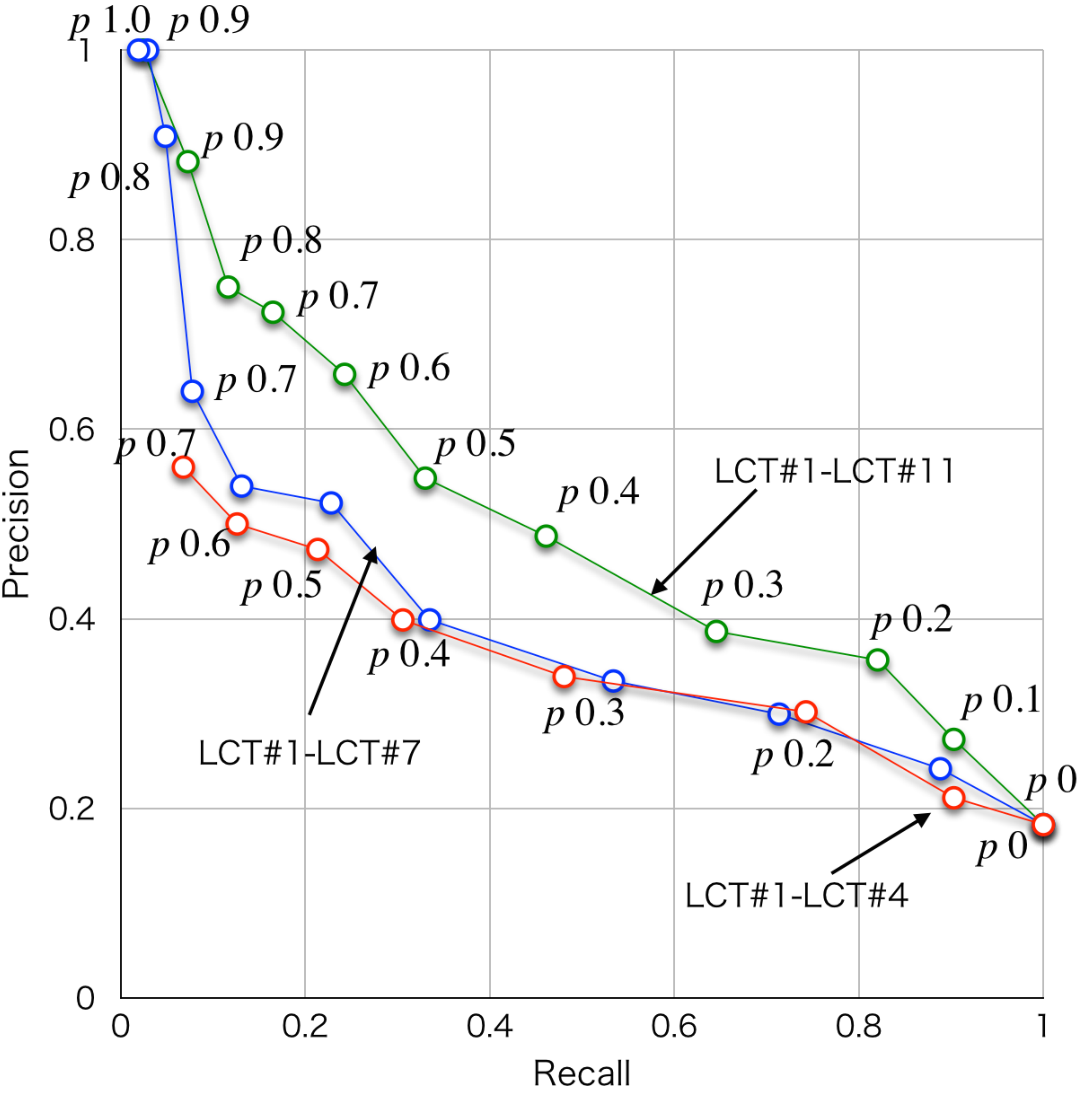}
\end{center}
\caption{Recall Precision curve (analysis basic in the first semester). 
Recall in the abscissa means 
the ratio of 
the number of actually failed students in the predicted failed students 
to 
the total number of actually failed students.
Precision in the ordinate means 
the ratio of 
the number of actually failed students in the predicted failed students 
to 
the total number of predicted failed students.}
\end{figure}


\begin{figure}[htbp]
\begin{center}
\includegraphics[height=6.5cm]{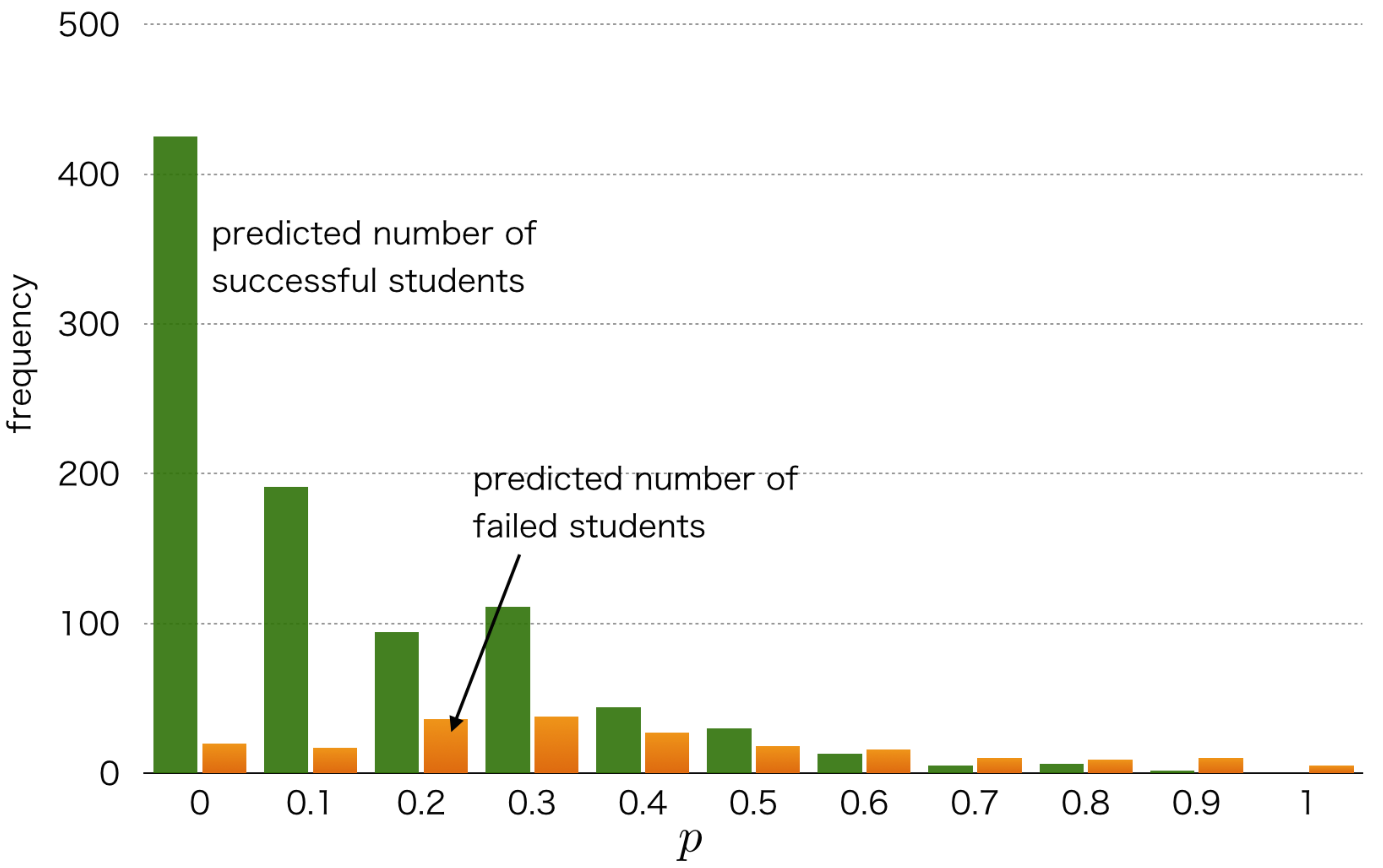}
\end{center}
\caption{Bar charts for predicted numbers of successful students and failed students to each $p$ using results from LCT no.1 to LCT no.11 (analysis basic in the first semester).}
\end{figure}


\section{Concluding Remarks}

Nowadays, it is crucial to identify students at risk for failing courses and/or dropping out as early as possible. By adopting the online testing systems such as the learning check testing, the LCT, for every class to check if students comprehend the contents of lectures or not, we can accumulate the information for learning analytics. This paper is aimed at obtaining effective learning strategies for students at risk by utilizing the learning analytics obtained from the online testings. 

To find students at risk as early as possible, we have proposed the newly developed method to identify successful/failed students by using the similarity of the trends of estimated students' abilities in the item response theory. 
The method uses the nearest neighbor methodology for determining the similarity of learning skill in the trends of estimated abilities.
In the cases of analysis basic subject in the first semester in 2017, the proposed method can point out almost half of the students failed in the final examination from the early stages. This result is superior to the hitting rate when we use the full data from the first to the last online testing results. 
We have applied ROC curve and recall precision curve to find the optimal threshold value for failure probability 
in investigating the accuracy of the proposed method precisely.



\section*{Acknowledgment}

The author would like to thank 
mathematical staffs at Hiroshima Institute of Technology.


\end{document}